\pdfoutput=1
\documentclass[runningheads]{llncs}

\usepackage[T1]{fontenc}
\usepackage{xurl}
\usepackage{hyperref}
\hypersetup{hidelinks}
\usepackage{tikz}
\usetikzlibrary{arrows.meta,positioning,fit,calc}
\usepackage{booktabs}
\usepackage{graphicx}
\graphicspath{{img/}}

\begin{document}

\title{Tiny Enough to Break In: Agentic Remote Access Trojans Powered by Small Language Models}

\titlerunning{Tiny Enough to Break In}

\author{Yuhan~You\inst{1} \and
Suhas~Adavelly\inst{1} \and
Victoria~Lovelace\inst{1} \and
Cameron~Berryman\inst{1} \and
Joel~Sadler\inst{1} \and
Daniel~Graham\inst{1}}

\authorrunning{Y. You et al.}

\institute{University of Virginia, Charlottesville, Virginia, United States}

\maketitle

\begin{abstract}
Agentic artificial intelligence raises a new security concern: cyber threats that reason, act, and adapt locally without continuous human direction. We examine this threat through an Agentic Remote Access Trojan (agentic RAT): a Remote Access Trojan augmented with a locally deployed Small Language Model (SLM). The SLM interprets host and network observations, selects actions, recovers from failed steps, and reduces reliance on an external operator. We implement the concept in a controlled, network-isolated lab built from Kali Linux, a Metasploitable2 target, LM Studio, and a local 8-billion-parameter Dolphin-family model. We then test whether a model this small can support autonomous cyber decision-making. This is architecturally feasible today. On commodity hardware, with no cloud service and no operator in the loop, the SLM closed the full observe-decide-act cycle: it interpreted ranked reconnaissance evidence supplied by the controller, selected actions, and obtained verified root-shell access on real vulnerable services. However, it is not yet operationally reliable. The same model hallucinated commands, misread output, and recovered from failure inconsistently, completing 10.9\% of a deliberately strict checklist. That gap reflects the limits of today's small models, not a ceiling on the concept. As SLMs improve, agentic endpoint systems may become more practical, more autonomous, and harder to detect, straining existing monitoring, containment, and policy-enforcement mechanisms. Real-world incidents in 2025--2026 already show AI-driven intrusions moving from concept toward practice. That makes the local, self-contained variant we study a plausible near-term direction, not a hypothetical one.

\end{abstract}

\section{Introduction}

Modern AI systems are no longer limited to passive text generation; given tools, memory, and an execution environment, they reason, plan, observe, and act. Defensive security has begun to absorb this shift through log analysis, vulnerability triage, and incident response. Offensive capability is evolving in parallel. Malware has long automated scanning, persistence, credential theft, and command execution, but not the judgment that directs them.

We define an agentic RAT as a remote-access architecture enhanced with an AI decision-making component that observes its environment, interprets evidence, selects actions, evaluates results, and revises its strategy with reduced human involvement. Its defining feature is not the presence of an AI model, but that model's embedding into a continuous observe-decide-act loop.

This framing matters because a conventional RAT already automates execution but not judgment. It lets an attacker remotely monitor and control a compromised system, yet the higher-level decisions remain a human operator's job: interpreting information from the target, choosing the next command, evaluating results, and correcting mistakes. An SLM inside that loop could absorb part of this role. Rather than waiting for each command, the implant would receive a broad objective and manage the intermediate steps itself. Existing work on AI-enabled cyber operations has largely focused on large language models, cloud-hosted agents, or AI-assisted penetration-testing tools. All assume continuous connectivity to an external service. A locally deployed SLM does not, which removes both a network dependency and a detection opportunity.

The premise is no longer purely hypothetical. In 2025 ESET reported PromptLock, a ransomware prototype that generates its attack scripts on the fly from a locally hosted open-weight model served through the Ollama API. It was later attributed to an academic proof-of-concept~\cite{eset_promptlock,ransomware3}. Google's Threat Intelligence Group catalogued malware families that query a language model at runtime, including PROMPTSTEAL, observed in active operations by a state-sponsored actor~\cite{gtig2025}. Anthropic disclosed a state-sponsored espionage campaign in which an AI agent executed an estimated 80--90\% of the operation with only sporadic human direction~\cite{anthropic_espionage}. Independently, an autonomous AI agent breached the internal infrastructure of a major model repository during a security evaluation, chaining exploits across a multi-day intrusion~\cite{hf_incident}. These incidents still rely on cloud-scale models or external services. The architecture we study removes that last dependency, moving the decision component onto the endpoint itself.

We report an exploratory study of that idea. Our contributions are: (i) a threat model and reference architecture for an SLM-driven agentic RAT; (ii) a reproducible, network-isolated testbed, released as a Docker Compose lab, in which a local Dolphin-family model drives a real reconnaissance-to-exploitation loop against a vulnerable target; and (iii) a benchmark-style evaluation quantifying how far current small models fall short of operating that loop reliably, and which behaviors defenders can exploit.

We draw a deliberate distinction between two claims. The first is \emph{architectural feasibility}: whether a local SLM can close the observe-decide-act loop end to end without human intervention or cloud support. Our results support it. The second is \emph{operational capability}: whether the resulting agent is an effective attacker. Our results do not yet support it, but the trajectory does. The model interpreted the ranked reconnaissance evidence the controller gave it, selected tools, and revised its strategy after failures. But it also hallucinated commands, misread terminal output, repeated ineffective actions, and recovered from failure inconsistently, completing 10.9\% of attempted service-level tasks. The security significance lies in the first claim, not the second. The loop closes today. The only remaining barrier is the reasoning that would make it dangerous, and that is the component improving fastest.

\section{Background}

\subsection{Evolution of Malware Automation}

Malware has evolved from manually operated programs into increasingly automated systems. Early malware required direct attacker interaction, but later families automated propagation, execution, and command-and-control communication. Botnets marked an important step, letting attackers control large numbers of compromised machines through centralized or distributed infrastructure~\cite{zeidanloo2010}. Exploit kits contributed similarly by packaging vulnerability exploitation into reusable frameworks, lowering the manual effort needed to compromise systems~\cite{hopkins2017}.

Recent AI advances suggest even greater malware autonomy is possible. Language-model agents can reason and act by observing information, selecting tools, and adjusting based on results~\cite{yao2022}. Tool-connected models can decide when to invoke tools and fold results into later reasoning~\cite{schick2023}. This matters because cyber operations may grow less dependent on fixed scripts and more adaptive to the environment.

The broader industry is also moving toward on-device small language models. Google AI Edge has expanded on-device SLM support across Android, iOS, and web platforms, including multimodality, retrieval-augmented generation, and function calling~\cite{sherwood2025}. Research on mobile SLMs shows they can run on smartphones, though size, context length, latency, and reliability remain constraints~\cite{pham2024}. Beyond feasibility, recent work argues that small models are not merely a compromise but a good match for agentic workloads. The specialized, repetitive, narrowly scoped subtasks that dominate agentic systems can be handled sufficiently, and more economically, by SLMs than by frontier models~\cite{belcak2025}. Consumer prototyping toolkits followed the same trajectory earlier, steadily raising the floor of what a single non-expert could build~\cite{sadler_cananyone}. An open-weight SLM on an endpoint applies that same floor-raising to offensive capability. As local models improve, agentic behavior running directly on endpoint devices becomes more plausible.

\subsection{Remote Access Trojans and Agentic RATs}

A RAT lets an attacker remotely access and control a victim system without authorization, supporting command execution, file access, credential theft, surveillance, and persistence. MITRE ATT\&CK categorizes the relevant tactics~\cite{mitre}, and behavioral research shows that families such as njRAT can be characterized through endpoint and network telemetry~\cite{prasse2021}. The same mechanism is legitimate or malicious depending on authorization and intent: a remote administration tool is used with consent, while a RAT is concealed from the user. We treat the agentic RAT as a working research definition rather than a standardized malware category. A traditional RAT gives an attacker visibility and control. An agentic RAT adds autonomous or semi-autonomous decision-making, drawing on the planning, memory, reflection, and tool use that characterize LLM agents~\cite{huang2024}.

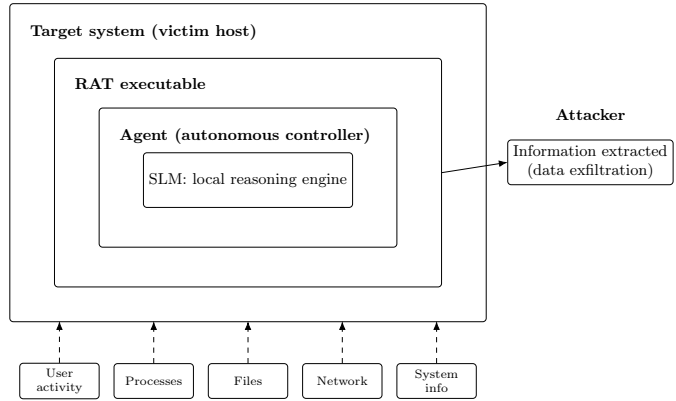
\begin{figure}[tb]
\centering
\resizebox{0.72\textwidth}{!}{%
\begin{tikzpicture}[
    font=\small,
    box/.style={draw, rounded corners=2pt, align=left},
]
  \node[box, minimum width=9.6cm, minimum height=6.4cm, anchor=north] (target) at (0,0) {};
  \node[font=\bfseries, anchor=north west] at ([xshift=3mm,yshift=-3mm]target.north west) {Target system (victim host)};

  \node[box, minimum width=7.8cm, minimum height=4.6cm, anchor=north] (rat) at ([yshift=-1.1cm]target.north) {};
  \node[font=\bfseries, anchor=north west] at ([xshift=3mm,yshift=-3mm]rat.north west) {RAT executable};

  \node[box, minimum width=6.0cm, minimum height=2.8cm, anchor=north] (agent) at ([yshift=-1.0cm]rat.north) {};
  \node[font=\bfseries, anchor=north west] at ([xshift=3mm,yshift=-3mm]agent.north west) {Agent (autonomous controller)};

  \node[box, align=center, minimum width=4.2cm, minimum height=1.1cm, anchor=north] (slm) at ([yshift=-0.9cm]agent.north) {SLM: local reasoning engine};

  \node[box, align=center, minimum width=1.6cm, minimum height=0.7cm, font=\scriptsize] (in1) at (-3.8,-7.6) {User\\activity};
  \node[box, align=center, minimum width=1.6cm, minimum height=0.7cm, font=\scriptsize] (in2) at (-1.9,-7.6) {Processes};
  \node[box, align=center, minimum width=1.6cm, minimum height=0.7cm, font=\scriptsize] (in3) at (0,-7.6) {Files};
  \node[box, align=center, minimum width=1.6cm, minimum height=0.7cm, font=\scriptsize] (in4) at (1.9,-7.6) {Network};
  \node[box, align=center, minimum width=1.6cm, minimum height=0.7cm, font=\scriptsize] (in5) at (3.8,-7.6) {System\\info};

  \draw[-Latex, dashed] (in1.north) -- (-3.8,-6.4);
  \draw[-Latex, dashed] (in2.north) -- (-1.9,-6.4);
  \draw[-Latex, dashed] (in3.north) -- (0,-6.4);
  \draw[-Latex, dashed] (in4.north) -- (1.9,-6.4);
  \draw[-Latex, dashed] (in5.north) -- (3.8,-6.4);

  \node[box, align=center, minimum width=2.8cm, minimum height=0.9cm] (exfil) at (6.9,-3.2) {Information extracted\\(data exfiltration)};
  \draw[-Latex] (rat.east) -- (exfil.west);
  \node[font=\bfseries] at ($(exfil.north)+(0,0.5)$) {Attacker};
\end{tikzpicture}%
}
\caption{Anticipated architecture of an Agentic Remote Access Trojan: a forward-looking reference design, not an artifact we built. A locally deployed SLM supports an autonomous controller embedded within a RAT executable, using host observations to inform decisions and potentially return selected information to an external operator.}
\label{fig:arat-architecture}
\end{figure}

Figure~\ref{fig:arat-architecture} illustrates this threat model. The local SLM is nested within an autonomous controller and RAT executable on the target system: host observations inform local reasoning and action selection, and selected information may be returned to an external operator. The diagram is a forward-looking reference architecture: our prediction of how such implants are likely to be structured as small models mature. It is not malware built or deployed in the controlled experiment.

Current small language models, however, have major limitations: they may hallucinate commands, misuse tools, forget context, or repeat ineffective strategies. Studies of on-device SLM deployment confirm mobile models can be useful but still face output-format failures, latency limits, context degradation, and instability~\cite{oliveira2026}. Agentic RATs should therefore be understood as an emerging risk area, not a mature malware class.

\subsection{Agentic Artificial Intelligence}

Agentic AI refers to systems that pursue goals through multi-step behavior: planning, tool use, memory, self-correction, and observation-action feedback loops. ReAct exemplifies this pattern by interleaving reasoning traces with actions~\cite{yao2022}. Reflexion takes a related approach: agents reflect on feedback from earlier attempts and store it in memory, improving later decisions without updating model weights~\cite{shinn2023}.

This tool access is what makes agentic AI powerful, and risky: a text-only model has limited direct impact, while a tool-connected model can affect real systems. Agentic AI expands the cybersecurity attack surface because agents retrieve information, invoke tools, rely on untrusted context, and execute actions through runtime supply chains~\cite{jiang2026}. Work on indirect prompt injection argues that agentic systems need system-level defenses to constrain what agents can observe, decide, and execute~\cite{xiang2026}. Offensive capability has also been measured directly. LLM agents have autonomously hacked websites and exploited one-day vulnerabilities from a CVE description alone~\cite{fang_websites,fang_oneday}. Benchmarks such as Cybench and CyberSecEval quantify how capability on capture-the-flag and exploitation tasks scales with model strength~\cite{zhang2024,cyberseceval}. These studies use large frontier models with cloud access. Our question is how much of that survives compression to a local endpoint model.

\subsection{LLM-Driven Offensive Security}
\label{sec:related}

Applying language models to offensive security is now an established research area. A recent survey catalogues 81 systems published between 2023 and 2026~\cite{he2026}. The closest precedent is Happe and Cito's closed loop between an LLM and a vulnerable VM over SSH, in which the model analyzes machine state, proposes attack vectors, and has them executed automatically~\cite{happe2023}. PentestGPT decomposes a long engagement across interacting modules to counter context loss~\cite{deng2023}. AutoAttacker targets the post-breach, ``hands-on-keyboard'' stage a RAT operator would otherwise perform manually~\cite{xu2024}. Most directly, Hackphyr fine-tunes a local 7B agent for network-security environments that performs comparably to much larger commercial models~\cite{rigaki2024}. Local deployment for offensive agents is therefore not itself novel. Our study differs in three respects. First, we use an \emph{off-the-shelf} instruction-tuned model rather than task-specific fine-tuning, to establish what is achievable without it. Second, our agent issues \emph{real} commands and parses genuine terminal output rather than evaluating in simulation, exposing failure modes (unsupported flags, interactive prompts, malformed syntax) a simulator abstracts away. Third, prior work treats the agent as an external attack tool, whereas we treat the decision loop as running inside the implant on the victim host. Together these choices measure a lower bound, not a best case: what an off-the-shelf model achieves when wired to real tools, without the advantage of fine-tuning.

\section{Methodology}
\label{sec:methodology}

\subsection{Experimental Environment}

The experiment was conducted in a controlled, network-isolated lab using only authorized systems built for testing, to evaluate whether a small language model could support an agentic penetration-testing workflow. The goal was not to deploy malware, but to study how an AI-assisted agent reasoned over reconnaissance data, selected tools, executed commands, and responded to failures.

\paragraph{What the testbed emulates.}
The architecture in Fig.~\ref{fig:arat-architecture} places the SLM inside an implant on a victim host; the testbed does not. We built no implant, persistence, or covert channel. Instead, we isolate the component that makes an agentic RAT distinctive: the decision loop. We run the SLM and its controller on the operator side, giving them the same interface an implant would have: a stream of observations from a live environment, and the ability to execute commands and read results. What transfers is evidence about the loop itself: whether a small local model can turn raw observations into justified actions, notice failure, and replan. What does not is anything requiring an implant in place, such as evasion, persistence, on-victim resource cost, or real command-and-control (Sect.~\ref{sec:limitations}).

\begin{figure}[tb]
\centering
\includegraphics[width=0.66\textwidth]{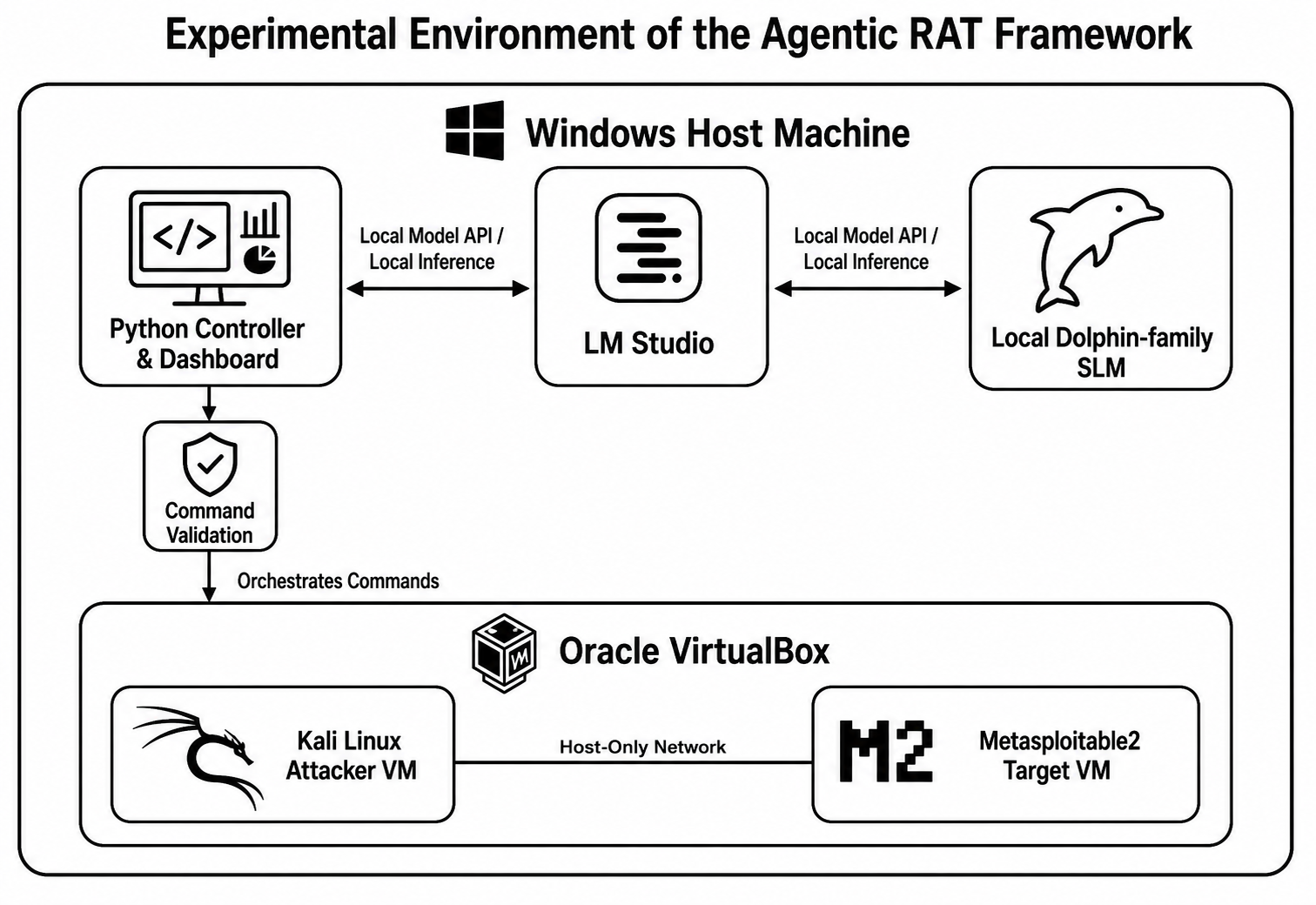}
\caption{Original VirtualBox-based experimental environment. The Windows host ran the Python controller, LM Studio, and the local Dolphin-family SLM, while an isolated host-only network connected the Kali Linux attacker VM to the Metasploitable2 target VM.}
\label{fig:virtualbox}
\end{figure}

An initial prototype ran the same components as two VirtualBox VMs, a Kali attacker and a Metasploitable2 target, on a host-only network. It resembled a conventional penetration-testing lab, but reproducing it depended on fragile VM, adapter, SSH, and tmux setup. Figure~\ref{fig:virtualbox} shows this original setup. We report results only from the containerized environment described below.

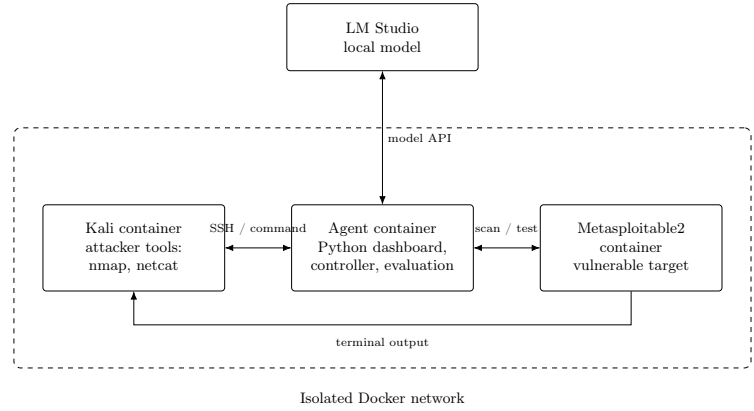
\begin{figure}[tb]
\centering
\resizebox{0.8\textwidth}{!}{%
\begin{tikzpicture}[
  font=\small,
  box/.style={draw, rounded corners=2pt, align=center},
]
\node[box, minimum width=4cm, minimum height=1.4cm] (lmstudio) at (0,4.4) {LM Studio\\\footnotesize local model};
\node[box, minimum width=3.8cm, minimum height=1.8cm] (kali) at (-5.2,0) {Kali container\\\footnotesize attacker tools:\\\footnotesize nmap, netcat};
\node[box, minimum width=3.8cm, minimum height=1.8cm] (agent) at (0,0) {Agent container\\\footnotesize Python dashboard,\\\footnotesize controller, evaluation};
\node[box, minimum width=3.8cm, minimum height=1.8cm] (meta) at (5.2,0) {Metasploitable2\\container\\\footnotesize vulnerable target};

\draw[Latex-Latex] (lmstudio) -- node[right, font=\scriptsize]{model API} (agent.north);
\draw[Latex-Latex] (kali) -- node[above, yshift=1.5mm, font=\scriptsize]{SSH / command} (agent);
\draw[Latex-Latex] (agent) -- node[above, yshift=1.5mm, font=\scriptsize]{scan / test} (meta);
\draw[-Latex] (meta.south) --++(0,-0.7) -| node[below, yshift=-1.5mm, pos=0.25, font=\scriptsize]{terminal output} (kali.south);

\node[draw, dashed, rounded corners, fit=(kali)(agent)(meta), inner ysep=16mm, inner xsep=6mm, label={[font=\small, yshift=-4mm]below:Isolated Docker network}] {};
\end{tikzpicture}%
}
\caption{Docker-based laboratory setup. The agent container communicated with the LM Studio model API, issued validated commands to the Kali container over SSH, and coordinated scanning and testing of the Metasploitable2 container within an isolated Docker network.}
\label{fig:docker-environment}
\end{figure}

To improve portability and reproducibility, we simplified the environment into a Docker Compose architecture with three coordinated containers: a Kali Linux attacker with OpenSSH and tmux pre-installed, the Metasploitable2 target, and an agent container hosting the Python control dashboard. The agent container connected to Kali over SSH and directed its reconnaissance and testing against Metasploitable2. All three communicated over an isolated Docker bridge network. This preserved the attacker-target-agent separation of the VirtualBox version while eliminating most manual VM and network configuration, making the environment easier to reproduce across host systems. Figure~\ref{fig:docker-environment} presents this containerized laboratory.

\begin{figure}[tb]
\centering
\includegraphics[width=0.82\textwidth]{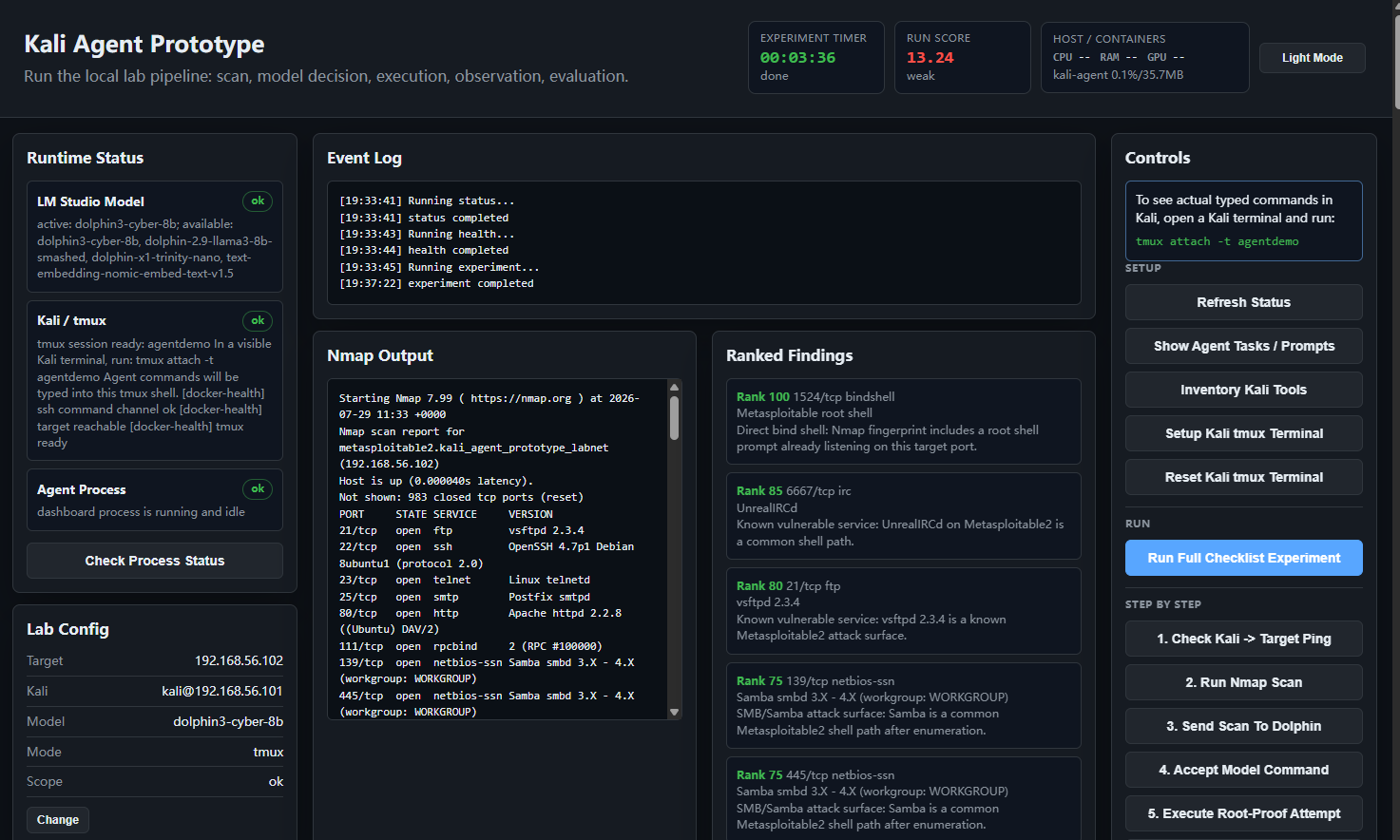}
\caption{The ARAT web dashboard during a checklist run: runtime status, ranked Nmap findings, the live event log, and operator controls to run or step through the autonomous loop. The headline run score (13.24/100) follows the fixed rubric of Sect.~\ref{sec:results}, where 0 is no verified access and 100 a fully completed checklist.}
\label{fig:dashboard}
\end{figure}

During execution, the agent communicated with the locally hosted model through LM Studio's OpenAI-compatible API. The model received structured reconnaissance evidence, including the controller's ranking of the discovered services, and generated command recommendations. The dashboard exposed connectivity checks for LM Studio, the Kali SSH service, the tmux session, and the target container, which we used to confirm the environment was ready before starting a run. The controller executed model-selected commands through Kali, captured terminal output, classified the result, and generated evaluation reports. Figure~\ref{fig:dashboard} shows the web dashboard used to launch and monitor these runs. It also let a researcher step through the loop one phase at a time, to inspect the model's reasoning before an action executed.

\subsection{Agentic Workflow}

\begin{figure}[tb]
\centering
\resizebox{0.82\textwidth}{!}{%
\begin{tikzpicture}[
  font=\scriptsize,
  box/.style={draw, rounded corners=2pt, minimum width=1.7cm, minimum height=1.4cm, align=center, font=\scriptsize},
  node distance=4mm
]
\node[box] (n1) {\textbf{1. Discover}\\scan target};
\node[box, right=of n1] (n2) {\textbf{2. Build context}\\parse evidence};
\node[box, right=of n2] (n3) {\textbf{3. Reason}\\propose action};
\node[box, right=of n3] (n4) {\textbf{4. Validate}\\check scope};
\node[box, right=of n4] (n5) {\textbf{5. Execute}\\run action};
\node[box, right=of n5] (n6) {\textbf{6. Observe}\\update state};
\node[box, right=of n6] (n7) {\textbf{7. Check}\\goal verified?};
\node[box, right=of n7] (n8) {\textbf{8. Terminate}\\record outcome};

\draw[-Latex] (n1)--(n2);
\draw[-Latex] (n2)--(n3);
\draw[-Latex] (n3)--(n4);
\draw[-Latex] (n4)--(n5);
\draw[-Latex] (n5)--(n6);
\draw[-Latex] (n6)--(n7);
\draw[-Latex] (n7)--(n8);

\draw[-Latex] (n7.south) -- ++(0,-0.7) -| (n3.south);
\node[font=\scriptsize] at ($(n5.south)+(0,-1.1)$) {Autonomous loop (bounded attempts)};
\end{tikzpicture}%
}
\caption{Agentic workflow implemented in the ARAT prototype. If the goal is not verified after an action, the updated observation is returned to the reasoning stage for replanning, up to a bounded number of attempts per checklist item.}
\label{fig:arat-workflow}
\end{figure}
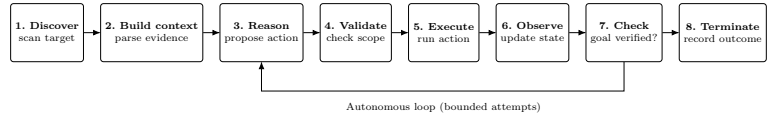

The prototype followed a ReAct-inspired workflow, combining environmental observation, SLM-based reasoning, tool execution, and feedback-driven adaptation. It added cybersecurity controls: command validation, execution logging, and explicit termination conditions. The workflow has three phases. Discovery and context generation prepare the evidence supplied to the SLM. Reasoning, validation, execution, observation, and goal checking form the bounded autonomous loop. Termination records the outcome once the goal is verified or the attempt limit is reached. Figure~\ref{fig:arat-workflow} summarizes these phases.

\subsubsection{Environment Preparation Phase.}

\paragraph{Discovery.}
From the Kali container, the system ran an authorized Nmap network and service scan against Metasploitable2, yielding open TCP ports, service names, version strings, and OS indicators.

\paragraph{Context Generation.}
The controller parsed this output into a structured per-service record: port, protocol, name, version, and a relevance note. It also ranked attack surfaces to reduce irrelevant context, giving a bind shell higher priority than general-purpose services such as HTTP, FTP, SSH, and databases. This ranking guided the model's attention but did not determine its action choice, a distinction that matters when interpreting our results (Sect.~\ref{sec:limitations}).

\subsubsection{Autonomous Agentic Loop Phase.}

\paragraph{SLM Reasoning.}
The structured evidence, current objective, previous actions, and observations were sent to the local model through LM Studio. The model was expected to identify a relevant service, select an appropriate tool or command, and briefly justify its choice. Each response thus paired an action proposal with a reviewable reasoning summary.

\paragraph{Validation.}
The controller checked each generated command for approved scope, correct target IP, and service-context match. It blocked common failure patterns: listener mode where a client connection was required, unsupported Netcat flags, ports absent from the scan, or tools mismatched to the active service.

\paragraph{Execution.}
Validated commands executed inside Kali. A web dashboard communicated over SSH and could mirror execution into a tmux session. The researcher could watch in real time or step through one phase at a time, without modifying the agent's actions. Each execution logged the proposed command, validation result, executed command, timing, stdout, stderr, exit status, and timeout status.

\paragraph{Observation.}
The controller converted each result into an updated observation: raw output, execution status, success/failure indicators, and a revised environment state. It searched for shell-session indicators and output from \texttt{whoami}, \texttt{id}, and \texttt{hostname}. It classified failures as syntax errors, missing tools, timeouts, connection or authentication failures, incompatible service responses, or insufficient evidence of completion.

\paragraph{Goal Check and Replanning.}
If the objective was not verified, the latest observation, previous action, validation feedback, and remaining attempt count returned to the SLM for another cycle. The controller tracked repeated command families to prevent unproductive loops, instructing the model to switch tool family, revise strategy, or declare no valid action remaining. The loop was capped at a bounded number of attempts per checklist item.

\subsubsection{Termination and Evaluation Phase.}
The workflow ended when the objective was verified or the cycle limit was reached. It recorded outcome, iteration count, proposed actions, validation decisions, executed commands, observations, and termination reason, tracking successful, unsuccessful, and inconclusive outcomes separately. This separates the SLM's raw reasoning from the controller-assisted agent's performance, in particular whether the model turned a failed action into a better next one.

\subsection{Evaluation Metrics}

We scored each run on a 100-point scale of three measured quantities: \emph{checklist completion} (60\%), the percentage of items reaching verified shell or root access; \emph{efficiency} (20\%), rewarding fewer attempts; and \emph{reliability} (20\%), penalizing dead-end loops and controller-blocked commands. The weighting reflects our judgment that verified access is what matters. It is not externally validated, so the score is meaningful only for comparing runs within this study. An established benchmark such as Cybench~\cite{zhang2024} would support cross-system comparison and is preferable in future work. We report the underlying task counts throughout, so results remain interpretable independently of the score.

The unit of work is the task: each open service or ranked attack surface from the Nmap scan counts as one. Success requires verifiable evidence of shell- or root-level execution: a shell prompt, or output from \texttt{whoami}, \texttt{id}, or \texttt{hostname}. Enumeration-only progress, however informative, was logged but never counted. The reported success rate is therefore a lower bound on the agent's usefulness to a human operator.

Two further behaviors matter for interpreting the score but resist direct measurement, so we treat them as qualitative categories over the execution logs. \emph{Planning quality} asks whether a chosen action followed from the evidence. A strong plan uses the correct target, a tool appropriate to the service, and a sensible progression from enumeration to exploitation. A weak one shows tool-service mismatch, invented assumptions, unsupported syntax, or repetition of an already-failed command family. \emph{Error recovery} asks whether the model changed strategy after a failure rather than reissuing the same command reworded. Both surface indirectly in the reliability term.

We also instrumented latency and resource use, capturing best-effort CPU, RAM, and GPU snapshots around each run. These are diagnostics rather than results: the containerized dashboard could not see the host clearly enough to make them comparable (Sect.~\ref{sec:results}). Autonomy proved hardest to quantify. We recorded blocked attempts, dead ends, and replanning events but never automated a human-intervention count, so we report it descriptively. Every run logged experiment- and service-level outcomes, model responses, command execution records, failure categories, and timing data for reproducibility.

\section{Results}
\label{sec:results}

\subsection{Environment Summary}

The experiment used the Docker-based lab with dolphin3-cyber-8b served via LM Studio, target IP 192.168.56.102, and Kali as the attacker.

We recorded three checklist experiments after the parser and controller updates. Across these runs the agent attempted 55 service-level tasks and completed 6, an overall success rate of 10.91\%. The remaining 49 ended in failure or in enumeration that never reached verified access. The average evaluation score was 12.72 out of 100, weak by the scale defined above, with an average runtime of 251.16 seconds. Two paths succeeded consistently: the 1524/tcp bind shell (root-shell proof) and 21/tcp FTP/vsftpd. Every other service remained incomplete in every run (Table~\ref{tab:results}).

The consistency of that outcome is itself the most informative result. The agent did not fail randomly: it succeeded on the same two services each time and failed on the same remainder each time. Both successes share a property: the scan evidence pointed to a single unambiguous action, with no multi-step reasoning between observation and verified access. The 49 failures are where a correct next step had to be inferred rather than read off.

\begin{table}[t]
\centering
\caption{Per-run results from three Docker-based checklist experiments using dolphin3-cyber-8b against Metasploitable2 (55 service-level tasks across the three runs). The two consistently successful paths were the 1524/tcp bind shell (rank 100) and 21/tcp FTP/vsftpd (rank 80), each verified by root- or shell-level command output.}
\label{tab:results}
\begin{tabular}{lrrr}
\toprule
Run & Score (/100) & Runtime (s) & Success rate \\
\midrule
1 & 11.67 & 259.73 & 9.52\% \\
2 & 13.24 & 252.98 & 11.76\% \\
3 & 13.24 & 240.78 & 11.76\% \\
\midrule
\textbf{Average} & \textbf{12.72} & \textbf{251.16} & \textbf{10.91\%} \\
\bottomrule
\end{tabular}
\end{table}

\subsection{Agent Performance}

Reconnaissance was the system's strongest component. Nmap reliably identified the target and discovered multiple open services, and the controller parsed this into structured findings (ports, service names, versions, and ranked attack surfaces), reducing the model's need to infer the environment. The agent was most reliable when scan evidence directly indicated an obvious shell path. The 1524/tcp service, ranked highest as a bind shell, produced a \texttt{root@metasploitable:/\#} prompt in the successful run. That is strong evidence that the agent uses scan evidence effectively when the required action is simple and direct.

Decision quality was poorer on more complex services: the model often recognized a service's importance but failed to choose the correct service-specific action. IRC, SMB, rsh/rlogin, SSH, Telnet, MySQL, PostgreSQL, Tomcat, and other web services were not completed. Common failures included repeated command families, incorrect tool syntax, tool-to-service mismatches, and commands requiring interactive input. Recovery was also limited. The controller detected repeated failures and prevented infinite loops, but the model often failed to generate a meaningfully different strategy after feedback, reaching dead ends by repeating similar commands or failing to switch tool families.

Resource measurement was partial. In Docker mode the dashboard recorded agent-container usage of about 0.1\% CPU / 34.91 MB RAM to 0.31\% CPU / 46.56 MB RAM. But host-level CPU, GPU, and RAM measurement was limited by the containerized dashboard's restricted visibility into the Windows host.

Overall, the agent completed limited shell-access tasks in a controlled lab, but autonomous penetration-testing capability remains weak. Reconnaissance and structured parsing worked well. The local Dolphin model struggled with operational command selection, replanning, and consistent tool use. The main limitation is not vulnerability discovery but controlling autonomous reasoning after the scan completes.

\section{Discussion}

\subsection{What the Loop Demonstrated}

The architectural claim holds. A model small enough to run on commodity hardware, given structured observations and a tool interface, sustained the full observe-decide-act cycle without human input. It read the controller's ranked reconnaissance evidence, proposed context-appropriate commands, consumed the output, and replanned, with no cloud service. The agent did not treat all 55 services as equivalent, but concentrated on the paths that evidence favored. We attribute the prioritization itself to the controller rather than to the model: the ranking is a hand-coded heuristic (Sect.~\ref{sec:limitations}), and what the model contributed was selecting and revising actions within it. Even so, choosing tools and commands from evidence about the environment is the behavior that distinguishes an agentic implant from a scripted one.

The qualification is that the loop closed reliably only where it had least work to do. Both repeatable successes were cases in which the scan output named the answer: a bind shell requiring a single connection. Only one inferential step separated observation from result. This tells us the plumbing works, not that the reasoning does.

\subsection{Where It Broke Down}

Everything requiring more than one inferential step degraded. The clearest deficit was long-horizon planning. The model could propose a plausible next step but could not hold a multi-stage workflow together, precisely what real operations demand across many observations, errors, and state changes. Failures clustered into recognizable modes: hallucinated commands and flags, tool-to-service mismatches, commands requiring interactive input the agent could not supply, and misreadings of terminal output that led it to treat failure as progress.

Recovery was the sharpest limitation and the most consequential for the threat model. The controller reliably detected repetition and prevented infinite loops. But detection is not recovery: told an approach had failed, the model frequently reissued a semantically identical command with altered wording rather than switching tool families. An agent that cannot convert failure into a different hypothesis cannot operate unattended for long, and limited context retention compounded this. We read these as properties of current small models rather than of the architecture. But that reading is an inference about model progress, not a result. What our data support is narrower: at present the reasoning component, not the surrounding machinery, is the binding constraint.

\subsection{Implications for Detection and Response}

An agentic implant would be harder to characterize than a conventional one. A traditional RAT exposes predefined functions and known command patterns, so its behavior is largely fixed. Our agent generated different command sequences across runs and varied tool selection with service context. The same property that frustrates signature matching, however, produced its inconsistency, and erratic behavior is itself a signal. The most useful defensive observation here is that the controller detected the agent's own failure modes cheaply. Dead-end loops, repeated command families, and blocked attempts were visible in execution telemetry. This suggests that monitoring focused on \emph{chains} of tool invocations is more promising than static signatures.

Two caveats bound these claims. First, the absence of command-and-control traffic follows from the architecture (local inference needs no external service), not from measured evasion. We deployed no implant and tested against no defensive product. Second, the recorded resource figures (roughly 0.1--0.31\% CPU and 35--47~MB RAM) describe the agent container, not a model on a victim host, where inference cost would dominate and be far more conspicuous. Treating either as evidence of stealth would overstate what we measured.

Incident response would also change in character. If an agentic component drives an intrusion, responders need to reconstruct not only which commands ran but why they were chosen and what evidence prompted them. Prompts, model outputs, tool calls, and validation decisions become forensic artifacts. This argues for building audit trails into any AI-assisted security system as a design requirement, not an afterthought. Our step-through dashboard (Fig.~\ref{fig:dashboard}) is an early instance, exposing the model's ranked findings, proposed action, and reasoning summary before each command executed. That design goal, giving an operator a visual view into how a composed system reaches its decisions, mirrors interface work on making the internals of modular prototyping systems visible to novices~\cite{sadler_bloctopus}. It also parallels the defensive turn in mechanistic interpretability, where humane interfaces onto a model's internal behavior are treated as a prerequisite for trusting or containing it. Building analysis on \emph{how} an agent decided, not only \emph{what} it did, is the more durable defensive posture.

\section{Limitations}
\label{sec:limitations}

Five limits bound these conclusions, beyond the implant-versus-testbed boundary described in Sect.~\ref{sec:methodology}. The first two are threats to validity that we consider serious enough to constrain the paper's central claims.

\paragraph{No controller-only ablation.}
We did not run the most informative control: executing the controller's ranked attack surfaces top-down with a rule-based tool mapping and no SLM. The ranking heuristic is hand-coded and assigns the 1524/tcp bind shell the highest priority, so that baseline would plausibly reproduce both repeatable successes. We therefore cannot show the model contributed positively to task completion, only that it did not prevent it. Our success figures should be read as an upper bound on what the SLM adds, which for two trivially reachable services may be nothing. This ablation is the first experiment we intend to run.

\paragraph{Possible training-data contamination.}
Metasploitable2 has been a standard teaching target since 2010, and step-by-step walkthroughs of its services are abundant online. A cyber-tuned model such as dolphin3-cyber-8b has plausibly encountered them in training. We cannot distinguish reasoning from recall on this target, so our two successes may reflect memorized procedure rather than inference over evidence. This weakens the successes more than it excuses the failures, and any claim about genuine reasoning requires a target absent from public write-ups.

The evaluation rests on a single, deliberately vulnerable target across three runs. Metasploitable2 is repeatable and safe to attack, but cannot represent the diversity of real networks. A target designed to be exploitable understates the reasoning burden a hardened host would impose, so our failure rate should be read as optimistic. The sample is also small enough that per-run differences in Table~\ref{tab:results} carry little weight. We draw conclusions only from the pattern that held across all three runs.

Model selection was constrained by what LM Studio supports within the host's limits, and we tested one model family. We therefore cannot separate limitations of small models in general from those of dolphin3-cyber-8b in particular, a distinction our central claim depends on. Finally, we tested short, supervised sessions rather than continuous operation, and ran no adversarial testing. Nothing here establishes whether the loop remains stable over long horizons, recovers from interruption, or withstands prompt injection, misleading telemetry, or manipulated scan results. The last is a natural defense against an agent that trusts its reconnaissance as completely as ours did.

\section{Broader Impacts and Ethical Considerations}

\paragraph{Conduct of this study.}
The work was performed entirely within a network-isolated lab, against Metasploitable2 (built and distributed to be attacked), on systems the authors controlled and were authorized to test. We built no implant, persistence mechanism, evasion capability, or command-and-control channel. The agent's actions were constrained throughout by a scope-checking validator and full execution logging. What we intend to release on acceptance is the testbed and evaluation harness, not a weaponized artifact (URL withheld during review). We judge publication net-beneficial. The architecture requires no capability an informed adversary lacks: local inference runtimes and vulnerability scanners are commodity tools. Defenders, meanwhile, have little empirical basis for reasoning about how such systems behave or fail. Our most transferable findings are in fact defensive, since the failure modes of Sect.~\ref{sec:results} are exactly the behavioral signals a monitor could exploit.

\paragraph{Why the concept matters despite weak results.}
An agentic RAT need not be local to be dangerous. The decision component could run on the compromised endpoint, or semi-locally on attacker infrastructure fed by victim telemetry. Either way, the operational change is the same: a system that interprets its environment, chooses next steps, and recovers from failure without an operator. RATs already expose files, credentials, browser sessions, messages, financial data, and device sensors. An agentic layer makes the resulting theft \emph{selective}, triaging what is worth taking rather than exfiltrating indiscriminately. The same logic applies at organizational scale, where one endpoint is a foothold and the expensive step is deciding where to go next.

\paragraph{Scale, not sophistication.}
The most plausible near-term harm follows from our results rather than in spite of them. Our agent succeeded roughly one time in ten. That would disqualify a human operator paid by the hour, but not a process that costs almost nothing to repeat. Unreliable automation still lowers the skill floor and raises attempt volume. Under-resourced targets (small businesses, schools, local governments, nonprofits) absorb the difference, being least likely to run monitoring that catches a clumsy agent. At national scale, the same capabilities bear on espionage and critical-infrastructure risk. None of this requires the reasoning gaps we measured to be solved, only that attempts become cheap.

\section{Future Work}

Error recovery was the weakest measured capability, and the one that most directly limited unattended operation. From a defender's vantage, it is the deficit most worth watching: it is the capability an adversary would need to improve first, and therefore the clearest early indicator that the threat is maturing. Two research directions probe it without waiting for better base models. The first is an explicit memory of attempted actions and their outcomes, so that failed strategies are excluded by construction rather than by the model remembering to avoid them. The second is decomposition of the loop into sandboxed, narrowly scoped sub-agents, each responsible for a single service or phase, rather than one model holding an entire operation together. Prior work on modular prototyping found that breaking a task into encapsulated building blocks measurably increased the quantity and novelty of what novices produced~\cite{sadler_buildingblocks}. The same logic predicts that modular SLM sub-agents will outperform a single monolithic controller on multi-stage operations. A scaling comparison across models available through LM Studio would establish whether recovery improves with parameter count or requires architectural support. A multi-host, heterogeneous testbed would extend the work to the scope-tracking behavior a single target cannot exercise.

We did not set out to build defenses, but our instrumentation points toward three complementary strategies. The first is \emph{behavioral telemetry}: the controller detected dead-end loops, repeated command families, and blocked attempts cheaply, so monitoring chains of tool invocations targets behavior an agentic implant cannot easily avoid. The second is \emph{richer sensing}: detecting an agentic loop is fundamentally a sensing problem, and finer-grained host and network telemetry surfaces anomalous decision patterns earlier. Prior work on modular sensor systems shows how much latent signal such instrumentation exposes~\cite{sadler_teamsense,graham_sonar}. The third is \emph{defense through education}: the most durable long-term defense is a workforce that understands these systems from the inside. Teaching practitioners to build and break a small model end to end is now cheap enough to be standard training. A public university course we drew on has students train a chat model from scratch and then attack it,\footnote{DS 6042, ``Machine Learning in Systems and Network Security,'' University of Virginia: \url{https://researcher111.github.io/ML-Security-Public/}.} and an open, browser-first adaptation lowers the barrier further.\footnote{\emph{Tiny AI: build a tiny ChatGPT from scratch:} \url{https://grassyhilltop.github.io/tiny-ai}.} Hands-on understanding of how a system is composed should precede attempts to secure it~\cite{sadler_anatomy}. That holds with more force now that anyone with an open-weight model can assemble an agent.

These directions also define concrete next steps. Because the controller already records decisions, tool usage, failures, and blocked actions, we can treat those traces as a detection dataset and test whether an agentic loop can be identified from execution telemetry alone. The same harness, connected to endpoint and authentication telemetry rather than an attack toolchain, would support the inverse application: an agent assisting investigation and triage, under guardrailed containment (scope checks, policy limits, approval, rollback).

\section{Conclusion}

We set out to separate two questions that discussions of AI-enabled malware tend to merge: whether a locally hosted small language model \emph{can} drive a remote-access architecture, and whether it can drive one \emph{well}. Our answers differ. A Dolphin-family model on commodity hardware sustained the full observe-decide-act loop, interpreting the controller's ranked reconnaissance evidence, issuing commands, reading results, and replanning, with no cloud service and no operator. It completed 10.9\% of attempted tasks, succeeding where a single inferential step separated observation from result, and failing systematically wherever recovery from error was required. A one-in-ten success rate reads as failure for a human operator paid by the hour. But it does not for a process that costs almost nothing to repeat across thousands of hosts, where a single verified intrusion can be consequential on its own.

The security-relevant finding is the gap between those two results, and where it sits. The surrounding machinery of parsing, ranking, validation, execution, and verification worked. The reasoning did not. That is an unfavorable place for a defender to find the bottleneck, because the machinery is engineering that will not regress, while the reasoning is the component improving fastest. We do not overstate this. That models will improve enough to close the gap is an expectation, not a measurement, and our single-target, short-horizon testbed cannot establish otherwise. What the study does establish is that the architecture is no longer the obstacle.

This cuts in a useful direction for defenders. The failure modes that made our agent ineffective (repeated command families, dead-end loops, and tool-service mismatches) were cheaply visible in execution telemetry. Monitoring built around chains of tool invocations, rather than static signatures, targets behavior an agentic implant cannot avoid producing. That window is open now and may not stay open.

\begin{credits}
\subsubsection{\ackname} The authors thank the University of Virginia for supporting this work.

\subsubsection{\discintname}
The authors have no competing interests to declare that are relevant to the content of this article.
\end{credits}

\end{document}